# Rayleigh Anomaly Induced Phase Gradients in Finite Nanoparticle Chains

Lior Michaeli[1-4,*], Ofer Doron[1-3], Yakir Hadad[1], Haim Suchowski[2,3], Tal Ellenbogen[1,3]

[1] *Department of Physical Electronics, Faculty of Engineering, Tel-Aviv University, Tel-Aviv 6779801, Israel*
[2] *Raymond and Beverly Sackler School of Physics & Astronomy, Tel-Aviv University, Tel-Aviv 6779801, Israel*
[3] *Center for Light-Matter Interaction, Tel-Aviv University, Tel-Aviv 6779801, Israel*
[4] Currently at: *Thomas J. Watson Laboratories of Applied Physics, California Institute of Technology, Pasadena, California, 91125, USA*

We report on the theoretical study of anomalous phase gradients induced by Rayleigh anomalies in finite nanoparticle chains. These phase gradients, defined with respect to the phase of the applied plane wave, cause a deviation of the diffraction directions from the chain relative to the direction expected from the grating equation for infinite chains. To study the effect theoretically, we use an analytical approach based on the discrete dipole approximation, which reveals the combinatorial nature of the multi-scattering process that governs the chain dynamics. We find an approximate closed-form solution to the particles' dipole moments by describing the single reciprocal system with a successive solution of two non-reciprocal, one-way systems. Within this framework, we obtain the chain excitation by means of interference between different scattering paths. Moreover, we show that the dipole moments along the chain are governed by recursive relations dictated by the generalized Fibonacci series. The presented results provide a new perspective for understanding nanoparticle arrays' dynamics. Specifically, the unique approach for analytically analyzing the spatial excitations of the array inclusions may shed new light on emerging applications of periodic traveling wave antennas in the optical regime, such as LIDARs, topological states analysis and arbitrary beam shaping schemes.

## I. INTRODUCTION

The interaction of waves with periodic structures stands at the heart of wave-matter dynamics. This interaction determines the electronic and photonic band structure of crystals, governs the operation principle of diffraction-grating spectrometers, and even enables the functionality of phased-array radars. Nanostructured periodic systems interacting with visible to infrared light have been the focus of extensive research over the past two decades, mainly due to advancements in nanofabrication techniques. A central topic of ongoing research in this field deals with the role of the electromagnetic interaction between the nanoparticles of an ordered array, and its influence on their excitation [1–3]. In the case of an infinite array illuminated by a plane wave, all nanoparticles will be excited by the same amplitude of dipole moment, with phase profile dictated according to the angle of incident light. However, for finite arrays, the interaction between the nanoparticles can cause spatial distribution of the dipole moment values over the array. The variety of phenomena emerging from this spatial distribution have been explored extensively in the microwave regime [4,5]. Over the past decade, they have gained increasing attention in the optical regime [6–15] due to their ability to control the functionality of various nanophotonic designs. The dipole moment variations may influence the far-field radiation from the array, for example, by causing a change in the diffracted beam width and shape or stimulating the radiation of leaky waves. Furthermore, they may reveal intriguing physical properties of the studied devices. For example, lately, it has been shown that via far-field analysis of leaky-waves radiation the topological properties and invariants of periodic arrays can be probed [16,17].

In periodic systems composed of resonant nanoparticles, the collective interaction supports a plethora of attractive and highly tailorable physical phenomena attributed to the enhanced light-matter interaction [1–3]. For example, the resonant collective interaction can be beneficial for nonlinearity enhancement and manipulations [18–21], induced transparency and slow light windows [22–24], sensing [25–28], lasing [11,29–31] and even for stimulating Bose-Einstein condensation at room temperature [32]. Recently, several important works studied how some of these phenomena are manifested also in finite-size arrays [11–15].

The dynamics in finite nanoparticle arrays are often explained by the modal solutions of the infinite array [8,9,33,34]. This description includes confined guided modes and expanded light cone modes [8,33], but also leaky modes, and continuous spectrum waves that are supported by the structure and may be excited also by global plane waves in the presence of edges [8]. While this approach is elegant and compact in the sense that it uses a few global wave functions that encapsulate the microscale interactions, the multi-scattering processes that build up the modes are hidden in such analysis.

Here, we explore the influence of the collective interaction in finite chains on the spatial distribution of the nanoparticles' excitation. Particularly, while the famous Rayleigh anomaly (RA) condition is associated with intensity anomalies of the light for over a century, we hereby show how non-trivial *phase anomalies* can emerge at that condition. In turn, these lead to a very interesting outcome, an *angular diffraction anomaly,* whereby the diffraction directions from finite chains deviates from the universally known grating equation. We show that this phenomenon is rooted in the multi-scattering process that governs the chain dynamics, and is inherent to chains with resonant nanoparticle. We develop a comprehensive theoretical model that captures the essential properties of this process and enable obtaining the chain excitation by means of interference between different scattering paths. The developed model is based on the discrete dipole approximation (DDA) in the spatial domain, which by dividing the original reciprocal system into two non-reciprocal, one-way systems, we obtain an insightful closed-form solution for the particles' dipole moments.

By the developed framework we obtain several new physical and mathematical insights on the system. We show that the chain's multi-scattering process can be described by an expression based on the multinomial coefficient, which reveals the combinatorial nature of the underlying dynamics. Moreover, we find an alternative mathematical solution to the system in terms of the recursive relations dictated by the generalized form of the Fibonacci series, a ubiquitous feature of nature. In addition, we use the developed model to explore the buildup of the studied mode and show how the scattering paths interference becomes interesting and non-trivial as the comprising particles become resonant. Finally, we link the newly developed theory of the microscopic picture, and the well-established theory of the macroscopic picture.

## II. THEORY

In order to analyze the spatial profile of the chain's excitation under plane wave illumination, we use the DDA [35,36]. This model serves to find the dipole moments' vector $p_i$ of each of the $N$ nanoparticles composing an array of arbitrary geometry ($i = 1, ..., N$) by solving a system of $3N$ coupled equations, which accounts for the mutual influence of all nanoparticles. To obtain physical insight, we will consider the simplified case of a finite 1D chain of $N$ identical, equally spaced nanoparticles. In addition, for simplicity, we will derive the model within the scalar approximation, where a specific polarization component governs the interaction. We denote the polarizability of the particle located at $r_i$, by $\alpha_{s,i}$, and express the dipole moments at the $i^{th}$ location as:

$$p_i = p(r_i) = \alpha_{s,i} E_{loc,i} \quad (1)$$

Where $E_{loc,i} = E_{loc}(r_i)$ is the local electric field at $r_i$, and stands for the field at the particle's location but in the absence of the particle itself. This field is composed of the applied field, denoted as $E_{app,i}$, and the retarded scattered fields from all other particles $E_{sca,i}$, at that location:

$$E_{loc,i} = E_{app,i} + E_{sca,i} = E_{app,i} + \sum_{j \neq i} G_{ij} p_j \quad (2)$$

Where $G_{ij}$ ($= G(|r_i - r_j|)$) is the electric dipole Green function that describes the interaction between the $i^{th}$ and $j^{th}$ dipoles [37]:

$$G_q = g_q e^{ikr_q}$$
$$g_q = \left[\frac{(1 - ikr_q)(3\cos^2(\theta_q) - 1)}{r_q^3} + \frac{k^2 \sin^2(\theta_q)}{r_q}\right] \quad (3)$$

Where $q = |i - j|$, $r_q = qd$ is the distance between particles $i$ and $j$, $d$ is the inter-particle spacing, $k = |k| = 2\pi n(\lambda)/\lambda$ is the wavenumber of the ambient medium and $\theta_q$ is the angle between the polarization of the incident field and the vector directed from $i$ to $j$. We will consider the case of $\theta_q = \pi/2$, i.e., where the polarization is perpendicular to the chain axis (transverse excitation). From Eqs. (1)-(3) a set of $N$ linear equations can be obtained:

$$\hat{A} p = E_{app} \quad (4)$$

Where $\hat{A}$ is the $N \times N$ interaction matrix, $p$ is $N \times 1$ dipole moment vector of the different particles and $E_{app}$ is $N \times 1$ vector of the applied electric field at each particle's location. The term $A_{ij}$ of the matrix $\hat{A}$ accounts for the interaction of particles $i$ and $j$ for $i \neq j$ (according to Eq. (3)), and for each particle's response to an external excitation for $i = j$ (according to Eq. (1)):

$$A_{ij} = \begin{cases} \alpha_s^{-1} & \text{for } i = j \\ -G_{|i-j|} & \text{for } i \neq j \end{cases} \quad (5)$$

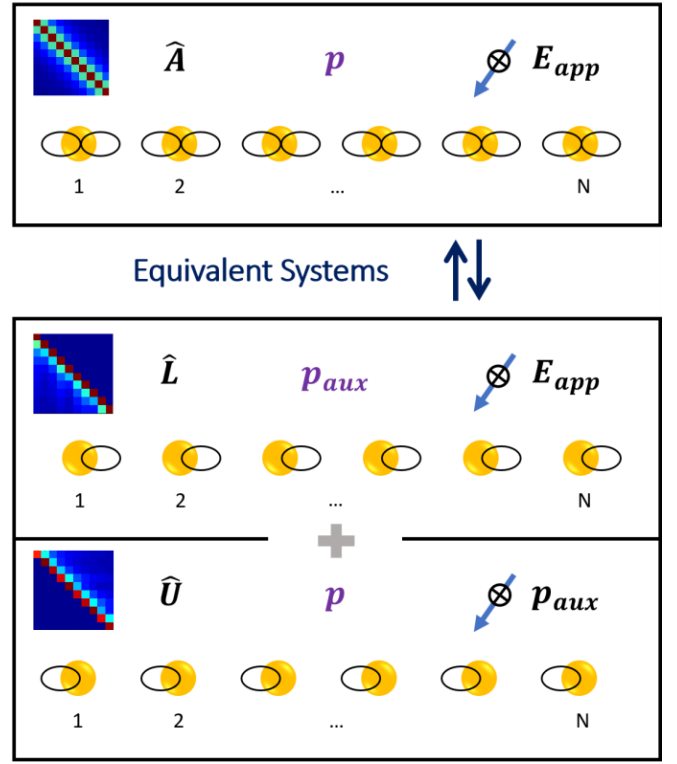

FIG. 1 Decomposition of a single reciprocal system described by the interaction matrix $\hat{A}$, to two artificial non-reciprocal, one-way systems described by $\hat{L}$ and $\hat{U}$. The top figure depicts the reciprocal system according to $\hat{A}$, where the particles scatter symmetrically, and $E_{app}$ is the applied field that serves to find $p$. The bottom figure shows the two non-reciprocal systems: First, excitation by $E_{app}$ and asymmetric scattering to the right, according to $\hat{L}$, serves to find $p_{aux}$. Then, excitation by $p_{aux}$ and asymmetric scattering to the left, according to $\hat{U}$, is used to find $p$. The black solid oblongs patterns that emanate from each particle depict its associated radiation pattern. The polarization of the applied field is perpendicular to the chain axis, as marked in the illustration.

The particles' dipole moments can be obtained from the inverse matrix $\hat{A}^{-1} \equiv \hat{B}$:

$$p = \hat{B} E_{app} \quad (6)$$

Typically, the matrix $\hat{A}$ is numerically inverted to obtain $\hat{B}$ and $p$.

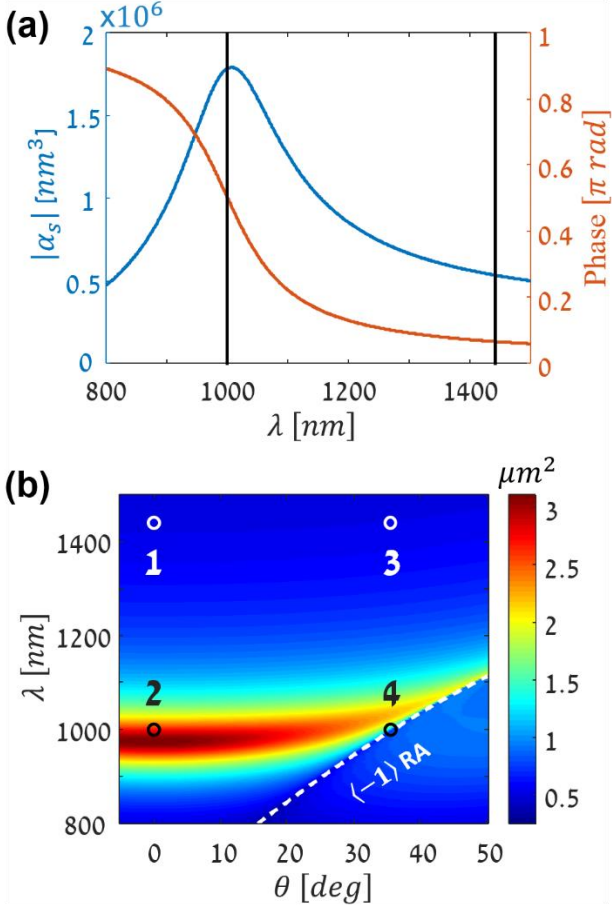

FIG. 2 The spectral response of the studied nanoparticle chain. (a) The amplitude (blue) and phase (orange) of the single nanoparticle polarizability. The black vertical lines denote the wavelengths in which the chain dynamics will be explored with the derived model. (b) The extinction per particle, vs wavelengths and incident angle, of an infinite chain with $d = 420\ nm$ and $n = 1.5$. These parameters are the same as for the finite chain studied within this paper. The dashed white line corresponds to the $\langle -1 \rangle$ RA. The chain will be explored at the four marked points in the figure: point 1 - off-resonance and off RA, point 2 – on resonance and off RA, point 3 - off-resonance and off RA, and point 4 – on resonance and on RA.

### A. The Derived Model

To obtain an insightful analytical solution for $p$, we choose to develop a model in which we *approximate* $\hat{A}$ by a different matrix, $\hat{A}_{Model}$, which can be analytically inversed. To that end, we first describe the single reciprocal system by two non-reciprocal systems, as described in FIG. 1. Explicitly, we perform an $LU$ decomposition to $\hat{A}$, such that[1]:

$$\hat{A} = \hat{L}\hat{U} \quad (7)$$

---

[1] We note that this kind of LU decomposition, without permutations, always exists provided that all the leading submatrices of $\hat{A}$ have a non-zero determinant [47].

Where $\hat{L}$ and $\hat{U}$ are lower and upper triangular matrices, respectively. Hence, the solution for $p$ can be obtained by successively solving the two following systems:

$$\hat{L}p_{aux} = E_{app}$$
$$\hat{U}p = p_{aux} \quad (8)$$

Where the auxiliary dipole moment vector $p_{aux}$, which is obtained from the solution to the first set of equations, serves as the excitation for the second set. These two sets of equations correspond to two artificial non-reciprocal, one-way systems, which can be described by particles with asymmetrical scattering patterns. As can be seen in FIG. 1, the first system characterized by interaction matrix $\hat{L}$, has only scattering towards the right side of the chain (solid black oblongs lines emanating from the particles), while the second system, described by $\hat{U}$, has only scattering towards the left side.

The derivation so far was exact. We proceed by taking advantage of the properties of $\hat{A}$ as a symmetrical Toeplitz matrix, i.e., $A_{ij} = A_{i+1,j+1} = A_{|i-j|}$, and perform simplifying approximation that holds for the cases of interest. The fact that $\hat{A}$ is a symmetrical Toeplitz matrix is attributed to the interaction between the particles, which depends only on their distances, i.e., $G(r_i - r_j) = G(|r_i - r_j|)$. We define a unit-less parameter that quantifies the strength of the nearest neighbors' interaction:

$$\zeta \equiv g_1 \alpha_s \quad (9)$$

Where $g_1$ is defined in Eq. (3). Then, for weakly interacting particles, i.e., for sufficiently small $|\zeta|$, $\hat{L}$ and $\hat{U}$ can be expressed by elements of $\hat{A}$ (see Section 1 of the Supplemental Material for an exact derivation of the validity condition). Specifically, if we decompose $\hat{A}$ to its lower and upper triangular parts as following:

$$A_{ij}^L = \begin{cases} A_{ij} & for\ i \geq j \\ 0 & for\ i < j \end{cases} \quad ;\quad A_{ij}^U = \left(A_{ij}^L\right)^T \quad (10)$$

Then, we can approximate $\hat{L}$ and $\hat{U}$ to be:

$$\hat{L} \rightarrow \hat{L}_{Model} \equiv \sqrt{\alpha_s}\hat{A}^L \quad \hat{U} \rightarrow \hat{U}_{Model} \equiv \sqrt{\alpha_s}\hat{A}^U \quad (11)$$

By solving the original system, according to the two steps shown in FIG. 1 and with the model matrices $\hat{L}_{Model}$ and $\hat{U}_{Model}$, the solution for the particles' dipole moments can be found. Going through the described procedure is equivalent to solving the original system in Eq. (4) with the model matrix $\hat{A}_{Model}$ instead of $\hat{A}$:

$$\hat{A} \rightarrow \hat{A}_{Model} \equiv \alpha_s \hat{A}^L \hat{A}^U \quad (12)$$

The validation of the derived model by direct comparison of the exact numerical solution and the approximated analytical solution according to the model assumption in Eq. (11) is presented in Section 2 of the Supplemental Material. The comparison is performed according to the cases described in the following sub-section.

## B. Studied Cases with the Derived Model

To apply the derived model to explore the RA-induced phase gradients phenomenon, we examine the extreme cases of very small and very large (maximal) polarizability, i.e., off and on resonance, respectively, and the cases of on and off RA condition. The maximal magnitude of the polarizability at wavelength $\lambda$ for a single mode, i.e., the dipole mode in our case, can be extracted from energy conservation considerations[2] [38]:

$$|\alpha_s^{Max}| = \frac{3}{2k^3} \quad (13)$$

Based on this, we examine a chain of identical nanoparticles with the following parameters: single nanoparticle polarizability with a Lorentzian form of $\alpha_s = A_0/(\omega_0^2 - \omega^2 + i\gamma\omega)$, amplitude $A_0 = 1 \times 10^{15}\ cm^3 s^{-2}$, resonance angular frequency $\omega_0 = 2\pi c/\lambda_0$, resonance free space wavelength $\lambda_0 = 1000\ nm$, angular frequency $\omega = 2\pi c/\lambda$, where $c$ is the speed of light, and damping constant $\gamma = 300\ THz$. The single nanoparticle polarizability is shown in FIG. 2(a). The vertical black lines denote the two wavelengths in which the chain dynamics will be explored with the derived model. In FIG. 2(b) we present the extinction cross section of a particle within an infinite chain, calculated according to [39]:

$$\sigma_{ext} = 4\pi k \cdot \Im(\alpha_{eff}) \quad (14)$$

Where $\Im$ denotes the imaginary part, and $\alpha_{eff}$ is the effective polarizability of a nanoparticle within an infinite chain [40]:

$$\alpha_{eff} = \left(\alpha_s^{-1} - S(k_\parallel)\right)^{-1} \quad (15)$$

Where $S(k_\parallel) = \sum_{q \neq 0} \hat{A}_q \cdot e^{-ik_\parallel \cdot r_j}$ is the array's incident angle dependent structural factor and $k_\parallel = k\sin(\theta)$ is the parallel component of the incident wave vector. FIG. 2(b) shows $\sigma_{ext}$ vs wavelength and incident angle, for an infinite chain with the exact parameters of the finite chain studied within the paper: ambient medium refractive index of $n = 1.5$, and the inter-particle spacing of $d = 420\ nm$. The dashed white line corresponds to the $m = -1$ RA, plotted according to the momentum conservation condition [41]:

$$k_\parallel + G_m = \pm k \quad (16)$$

Where $G_m = 2\pi m/d$ is a reciprocal lattice vector and $m$ denotes the order of the RA. The chain will be explored at the four marked points in the figure: point 1 - off-resonance and off RA ($\lambda = 1440\ nm, \theta = 0°$), point 2 – on resonance and off RA ($\lambda = 1000\ nm, \theta = 0°$), point 3 - off-resonance and off RA ($\lambda = 1440\ nm, \theta = 35.5°$), and point 4 – on resonance and on RA ($\lambda = 1000\ nm, \theta = 35.5°$). Polar plots of the scattered fields at these 4 points, for a finite array of 50 particles, are presented in section 3 of the Supplemental Material.

## C. Multi-scattering process

According to the above derivation, in order to solve for $p$ we need to solve two systems with triangular Toeplitz matrices. Each has an analytical solution for the required matrix inversion [42,43].

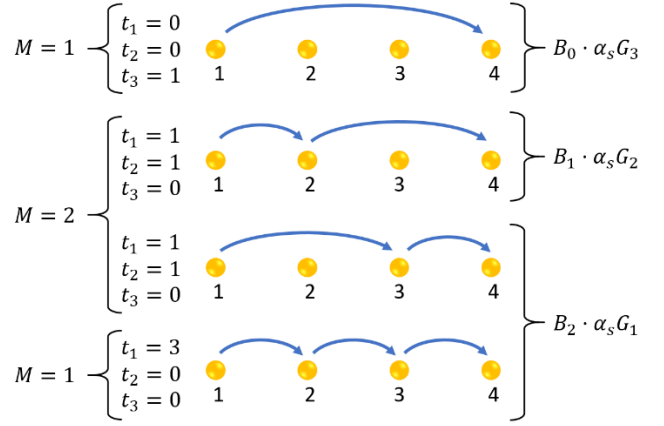

FIG. 3 Multi-scattering process. Counting all possible scattering paths from particle 1 to particle 4, while scattering only towards the right direction of the chain. The associated $t$'s and multinomial coefficients, $M$'s, are written for each scattering path on the left side. Interpretation in terms of recursive relations is given on the right side: The top bracket depicts only the direct scattering from the first ($B_0$) to the fourth particle ($\alpha_s G_3$). The next bracket shows the scattering from the first to the second particle ($B_1$) and then the resulting path from the second to the fourth particle ($\alpha_s G_2$). The last bracket marks the two contributions of scattering from the first to the third particle ($B_2$) and then the resulting path from the third to the fourth particle ($\alpha_s G_1$).

The inverse of $\hat{A}^L/\hat{A}^U$ is also a finite lower/upper triangular Toeplitz matrix [42]:

$$B_{ij}^L \equiv \left(A_{ij}^L\right)^{-1} = \begin{cases} B_{|i-j|} & for\ i \geq j \\ 0 & for\ i < j \end{cases} \quad (17)$$

$$B_{ij}^U = \left(B_{ij}^L\right)^T$$

Where $B_q = B_{|i-j|}$ $(q = |i-j|)$ is defined as:

$$B_q = \alpha_s \sum_{t_1 + 2t_2 + \cdots + qt_q = q} \binom{t_1 + \cdots + t_q}{t_1, \ldots, t_q} (\alpha_s G_1)^{t_1} \cdots (\alpha_s G_q)^{t_q} \quad (18)$$

And the multinomial coefficient is defined as:

$$M(\vec{t}) \equiv \binom{t_1 + \cdots + t_q}{t_1, \ldots, t_q} = \frac{(t_1 + \cdots + t_q)!}{t_1! t_2! \ldots t_q!} \quad (19)$$

Where $\vec{t}$ specifies the value of $t_1, \ldots, t_q$. The number of absorption-reemission processes for each $\vec{t}$ is $T \equiv \sum_i t_i$. As we explain in the following, the physical meaning of each of Eq's. (18) terms can be well understood. First, we see that $B_0 = \alpha_s$, which accounts for the particles' response to the applied field. Then, the expression for $B_q$, for $q \geq 1$, describes the scattering of a certain particle on some other particle, where the two particles are separated by a distance of $qd$. The sum in Eq. (18) runs over all sets of $(t_1, t_2, \ldots, t_q)$ that their weighted sum $\sum i t_i$ adds up to $q$. This sum describes all combinations of scattering paths between the two particles, in which light can interact with any particle in the chain while maintaining the rule of a unidirectional scattering. For example, looking at the case of 4-particle chain, as shown in FIG. 3, we see that there are four possible paths of scattering from particle 1 to particle 4: first $1 \to 4$, second $1 \to 2 \to 4$ third $1 \to 3 \to 4$ and fourth $1 \to 2 \to 3 \to 4$. The associated sum in $B_3$ has three terms corresponding to: first $t_1 = 0, t_2 = 0, t_3 = 1$, second $t_1 = 1, t_2 = 1, t_3 = 0$ and third $t_1 = 3, t_2 = 0, t_3 = 0$. The degeneracy of the two scattering paths $1 \to 2 \to 4$ and $1 \to 3 \to 4$, which both are described by $t_1 = 1, t_2 = 1, t_3 = 0$, is accounted for by the associated multinomial coefficient, $M = 2$. Generally, different paths which are composed of similar scattering segments but ordered differently, correspond to a single term in the sum, with weight according to the associated multinomial coefficient. In FIG. 4 we show how the number of scattering paths increases rapidly as a function of the number of particles.

---

[2] We keep the discussion general by referring to the fundamental property which determines the particles response, i.e. the polarizability, but we note that the values we analyze can be realized both with metallic and dielectric nanoparticles (see [38]).

To simplify the expression for $B_q$, we divide it to a complex amplitude term, $b_q$, and a phase term:

$$B_q = b_q e^{ikr_q} \quad (20)$$

The term $e^{ikr_q}$ describes the phase accumulation of a wave propagating a distance of $r_q = qd$. By aiming to find $b_q$ we can neglect the phase resulting from free-space propagation, which is especially beneficial around the RA condition. By substituting the relation $G_q = g_q e^{ikr_q}$ from Eq. (3) into Eq. (18) we note that the phase in each term of the sum in $B_q$ due to the multiplication $G_1^{t_1} G_2^{t_2} \ldots G_q^{t_q}$ will be $e^{ikd(t_1 + 2t_2 + \cdots + qt_q)} = e^{ikr_q}$. Therefore, we find $b_q$ to be:

$$b_q = \alpha_s \sum_{t_1 + 2t_2 + \cdots + qt_q = q} M(\vec{t}) (\alpha_s g_1)^{t_1} \cdots (\alpha_s g_q)^{t_q} \quad (21)$$

Importantly, according to Eq. (6), the solution for each $p_q$ is determined by the various elements of the matrix $\hat{B}$. Therefore, the underlying physical dynamics are embedded in the derived $b_q$ terms.

To examine the appearance of nontrivial phase gradients over the chain and better understand the meaning of the $B_q$ terms, it is insightful to examine the difference in dipole moment of two adjacent particles in the chain. By considering first the solution of the unidirectional system described by $\hat{L}$, and for the case of normal incidence, i.e., $E_{app,q} = E_0$ $(= 1 V/nm)$, we get:

$$\Delta p_{aux,q} \equiv p_{aux,q+1} - p_{aux,q} = B_q E_{app,q}^* \quad (22)$$

Where $E_{app,q}^*$ denotes the complex conjugate of $E_{app,q}$. We see that $B_q$ determines the difference in dipole moments of particles $q + 1$ and $q$. This result is expected since particle $q$ experiences scattering from $q - 1$ particles from its left (particles 1 to $q - 1$), much the same as particle $q + 1$ experiences scattering from $q - 1$ particles from its left (particles 2 to $q$). In addition, particle $q + 1$ experiences scattering from particle 1, described by $B_q$. We can generalize the relation in Eq. (22) for the case of oblique incidence, $E_{app,q} = E_0 e^{-ik_{||}r_q}$, by looking at the angle-normalized dipole-moment difference:

$$\Delta \tilde{p}_{aux,q} \equiv \tilde{p}_{aux,q+1} - \tilde{p}_{aux,q} = B_q E_{app,q}^* \quad (23)$$

Where:

$$\tilde{p}_{aux,q} \equiv \frac{p_{aux,q}}{e^{i\phi_q}} \quad (24)$$

and $\phi_q = arg(E_{app,q})$. By looking at $\tilde{p}_q$ and $\Delta \tilde{p}_q$ we eliminate the spatial phase profile in the chain attributed to the applied field and examine only the phase that arises due to the interaction between the nanoparticles.

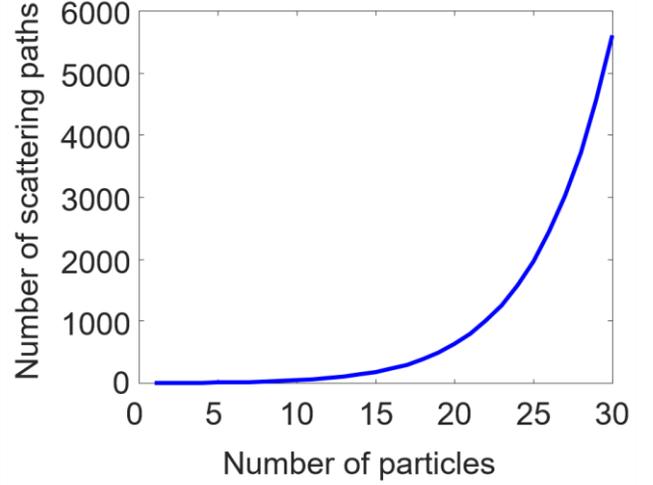

FIG. 4 Number of scattering paths as a function of the number of particles.

The second unidirectional system, described by $\hat{U}$, is solved in the same manner but for scattering in the opposite direction of the chain, according to FIG. 1. Thus, the equivalents of Eqs. (23)-(24) for that system are:

$$\Delta \tilde{p}_q \equiv \tilde{p}_{q+1} - \tilde{p}_q = -B_{N-q} p_{aux,q}^* \quad (25)$$

$$\tilde{p}_q \equiv \frac{p_q}{e^{i\phi_q}} \quad (26)$$

These quantities, i.e., the angle-normalized dipole moment $\tilde{p}_q$ and its difference $\Delta \tilde{p}_q$, play a crucial role in the understanding of the excitation evolution along the chain, as we further discuss in section III. Specifically, in the vicinity of a RA condition, it is convenient to describe the difference in dipole moment from Eqs. (23) and (25), by the complex amplitude $b_q$. In particular, the insightful relations $\Delta \tilde{p}_{aux,q} \approx b_q$ or $\Delta \tilde{p}_q \approx -b_{N-q}$ hold near a RA towards the right side (positive order RA) or towards the left side (negative order RA) of the chain, respectively.

### D. Generalized Fibonacci series

While the solution for the Triangular Toeplitz matrix in terms of the multinomial coefficient reveals essential properties of the multi-scattering processes that govern that chain dynamics, there is an additional important solution. This solution is based on the generalized Fibonacci polynomials [43], and it uncovers the recursive relations that underlay the excitation of the chain. The additional solution for $B_q$ has the following form:

$$B_q = \alpha_s F_{N-1,q}(\vec{W}) \quad (27)$$

Where $\vec{W} = (W_1, W_2, \ldots, W_{N-1}) = \alpha_s(G_1, G_2, \ldots, G_{N-1})$ is a weight vector that determines the contribution of each element in the recursive formula for $F_{N-1,q}(\vec{W})$:

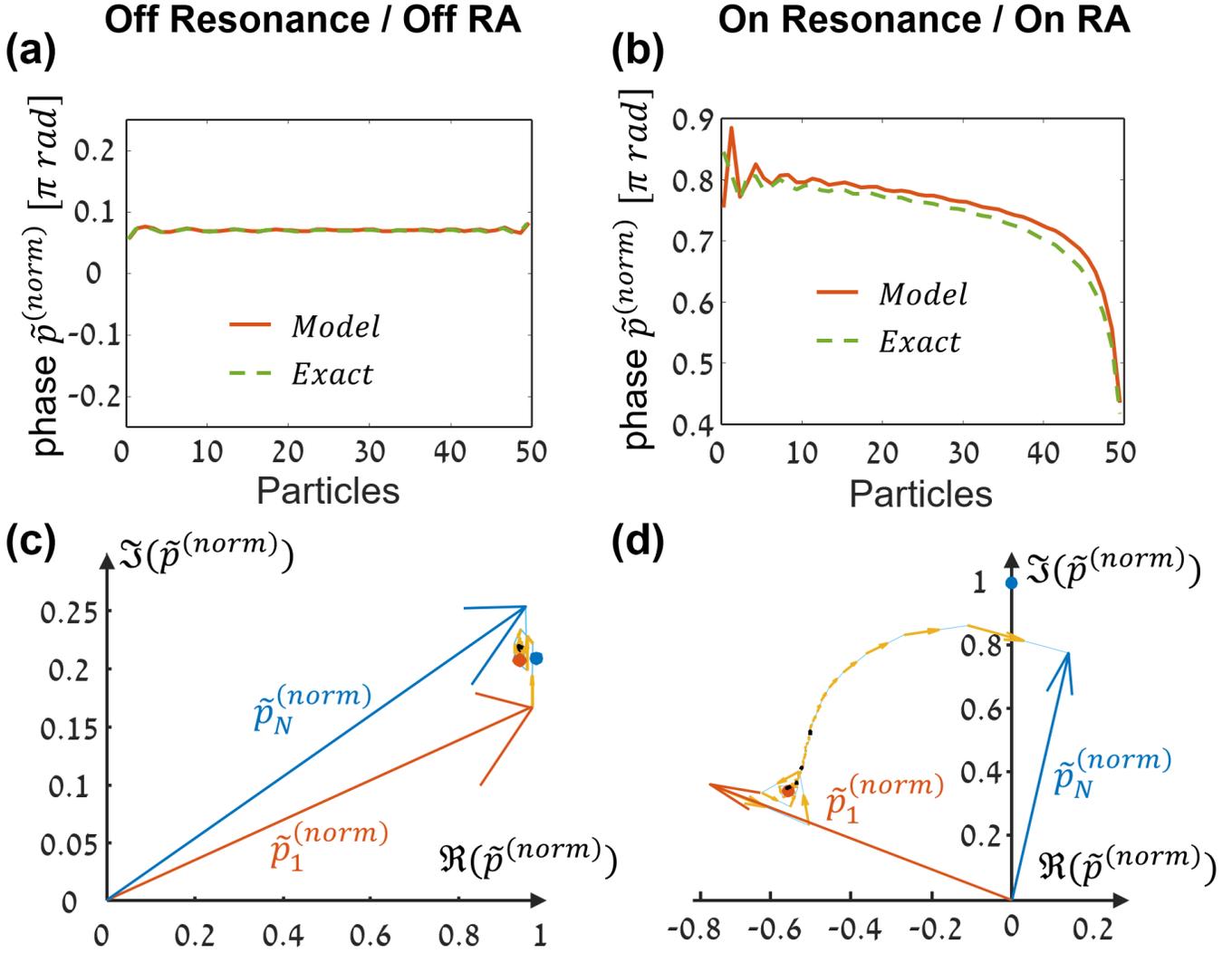

FIG. 5 RA induced phase gradient. (a) The phase of $\tilde{p}^{(norm)}_q$ for off-resonance excitation ($\lambda = 1440\ nm$), and far from a RA. (b) The phase of $\tilde{p}^{(norm)}_q$ for on-resonance excitation ($\lambda = 1000\ nm$) and in vicinity to a RA ($\lambda = 996\ nm$). In (b) the nearly monotonic phase trend that spans $\sim 0.5\pi\ rad$ can be seen. In (a) and (b) the dashed green and solid orange lines correspond to the exact solution and the solution based on the model in Eq. (12), correspondingly. The model follows the same trend as the exact solution. (c) The trajectory of $\tilde{p}^{(norm)}_q$ of all the nanoparticles in the chain, on the complex, for off-resonance excitation. (c) The trajectory of $\tilde{p}^{(norm)}_q$ for on-resonance excitation. In (c) and (d) the orange arrows point to $\tilde{p}^{(norm)}_1$, and the blue arrows to $\tilde{p}^{(norm)}_N$. In the middle between these two, the yellow arrows, with the light blue guiding lines, show the difference in $\tilde{p}^N_q$. The arrows of every tenth particle are marked in black ($N = 10, 20, 30, 40$). We note that the arrows in (d) point to the direction of particles 1 to 50, while the coherent interaction, according to the $m = -1$ RA, is in the direction of particles 50 to 1. In addition, the blue and orange points show the excitation of the single nanoparticle, and the excitation a nanoparticle within an infinite chain, respectively. The values of $\zeta$ are: Off resonance: $\zeta = 0.05 \cdot e^{0.19 i\pi}$, on resonance: $\zeta = 0.37 \cdot e^{0.58 i\pi}$.

$$F_{N-1,q}(\vec{W}) = \begin{cases} W_1 F_{N-1,q-1}(\vec{W}) + \cdots + W_{N-1} F_{N-1,q-N+1}(\vec{W})\ ,\ q > 0 \\ 1,\qquad q = 0 \\ 0,\qquad q < 0 \end{cases} \quad (28)$$

Where $F_{N-1,q}(\vec{W})$ are the generalized Fibonacci polynomials. In this recursive solution, each of the $B_q$ terms depend on all the previous $B_i$, where $i < q$. Similarly, we can define recursively the $b_q$ from Eq. (21) by:

$$b_q = \alpha_s F_{N-1,q}(\vec{w}) \quad (29)$$

Where $\vec{w} = \alpha_s(g_1, g_2, \ldots, g_{N-1})$. Equations (27) and (29) are equivalent to Eqs. (18) and (21), respectively. Specifically, the sum over the different scattering paths in Eqs. (18) and (21) is replaced by an equivalent recursive sum in Eqs. (27) and (29). The recursive interpretation for the sum can be understood from FIG. 3 According to Eqs. (27) and (28) we can find:

$$B_3 = B_0 \cdot \alpha_s G_3 + B_1 \cdot \alpha_s G_2 + B_2 \cdot \alpha_s G_1 \quad (30)$$

This describes the scattering, depicted in FIG. 3, from particle 1 to particle 4 as following: The first term – only the direct scattering from the first ($B_0$) to the fourth particle ($\alpha_s G_3$). The second term – the scattering from the first to the second particle ($B_1$) and then the resulting path from the second to the fourth particle ($\alpha_s G_2$). The third term – the scattering from the first to the third particle ($B_2$) and then the resulting path from the third to the fourth particle ($\alpha_s G_1$). This partition is described on the right side of FIG. 3. We note that the calculations of $B_q$ and $b_q$ are much more efficient using the recursive relation in Eqs. (27)-(29) thanks to the ability to perform memoization. Explicitly, the calculation can be optimized by storing the results of previously calculated $B_q$ and $b_q$ and avoid re-computation.

## III. RA INDUCED PHASE GRADIENTS

The derived theoretical model allows us not only to acquire a closed form solution for dipoles' excitations, but also to analyze the chain dynamics through a new perspective. By looking at the multi-scattering process as the source of the interparticle interactions, we can understand the evolution of edge effects along the chain by using properties intrinsic to the chain. Specifically, we use the theoretical model to examine how a unidirectional RA can induce phase gradients in finite nanoparticle chains. For that, we look at the same chain from FIG. 2, though for an oblique angle of incidence, such that a unidirectional RA condition will be fulfilled. Specifically, we illuminate the chain at $\theta = 35.5°$ and compare the dynamics at point 3 (1440 $nm$) and point 4 (1000 $nm$), from FIG. 2(b). For that angle of incidence, we can find the wavelength of the $m = -1$ RA to be $\lambda = 996\ nm$, using Eq. (16). In FIG. 5 we examine the phase of the normalized dipole moment $\tilde{p}_q^{(norm)} = \tilde{p}_q/|p_s|$ of all the nanoparticles, for the two described cases. In FIG. 5(a) we first show the case of off-resonance excitation ($\lambda = 1440\ nm$), and far from a RA (point 3). We can see the minute oscillations of the phase, which spans less than $0.03\pi\ rad$, and show no monotonic trend. On the contrary, in FIG. 5(b), we excite the chain at the localized resonance of the nanoparticles ($\lambda = 1000\ nm$) and in vicinity to a RA (point 4). We can see the nearly monotonic phase trend that spans $\sim 0.5\pi\ rad$. The phase gradually increases from the last to the first particle, according to the direction of the coherent scattering buildup for the $m = -1$ RA. To better examine the spatial phase profile and the entire chain excitation, we plot the trajectory of $\tilde{p}_q^{(norm)}$ of all the nanoparticles in the chain, on the complex plane (FIG. 5(c) and 5(d)). The orange arrows point to $\tilde{p}_1^{(norm)}$, and the blue arrows to $\tilde{p}_N^{(norm)}$. In the middle between these two, the yellow arrows, with the light blue guiding lines, show the differences in $\tilde{p}_q^{(norm)}$, such that the evolution of $\tilde{p}_q^{(norm)}$ of the entire chain can be tracked. In addition, the blue points in the figure correspond to the excitation of the single nanoparticle:

$$p_s^{(norm)} \equiv \frac{\alpha_s}{|\alpha_s|} = e^{i\phi_{\alpha_s}} \quad (31)$$

Where $\phi_{\alpha_s} = \arg(\alpha_s)$. This is equivalent to the excitation of a nanoparticle within a hypothetical chain with no interactions. The orange points in the figure correspond to the excitation of a nanoparticle within an infinite chain with the same parameters [40]:

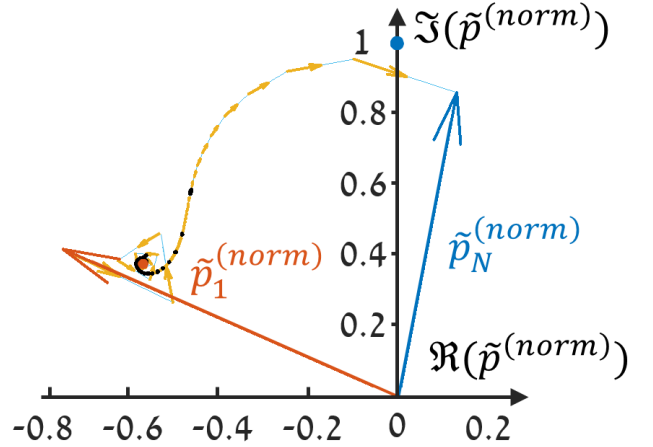

FIG. 6 RA induced phase gradient in a longer chain. The figure represents the complex plane trajectory for a chain with the same parameters as in FIG. 5(d), though for $N = 500$ particles. The exact same trend as in FIG. 5(d) can be seen. The additional particles, relative to the case of $N = 50$, were added mostly to the spiral at the beginning of the chain.

$$p_\infty^{(norm)} = \frac{\alpha_{eff}}{|\alpha_s|} \quad (32)$$

Where $\alpha_{eff}$ is defined in Eq. (15). In addition, to easily track the evolution of the chain as a function of the number of particles, we mark every 10 particles by a black arrow (particles 10,20,30,40 are marked).

We can see that in FIG. 5(c) the entire range of evolution on the complex plane is confined to a substantially smaller region than in FIG. 5(d). Moreover, in FIG. 5(c), the orange and blue points are very close, such that no substantial change of the single particle response occurs due to the interaction. On the other hand, the same two points in FIG. 5(d) are considerably separated. Intriguingly, in that case, we can see that the end of the chain responds similar to the single particle case, while the beginning of the chain responds similar to an infinite chain. This behavior is attributed to the coherent buildup of the $m = -1$ RA, as we further explain. The last particle of the chain experiences only the incoherent scattering from all the chain, which sums up to a minor change in its local field and, therefore in its excitation, relative to the single particle case. Explicitly, the incoherent scattering translates to $p_{aux,N} \approx E_{app,N}$, which in turn translates to $p_N \approx \alpha_s E_{app,N}$. Proceeding from that particle towards the beginning of the chain, each particle experiences a superposition of incoherent scattering from its left, and coherent scattering from its right. While the strength of the incoherent scattering does not change much from particle to particle, the coherent scattering changes significantly as more particles are added. This is what gives rise to the monotonic trend of the trajectory (particles 50 to 30), as further explained in the following. At some point (~ particle 30), adding more particles to the coherent scattering does not influence much the resulting amplitude of the scattered field. From that point to the beginning of the chain starts the observed spiral with minor variations in the excitations. In the very first particles of the chain (particles 1 to 5) the variations are larger. This is a consequence of the influence of each particle that is added to the incoherent scattering being more significant where there are only a few particles.

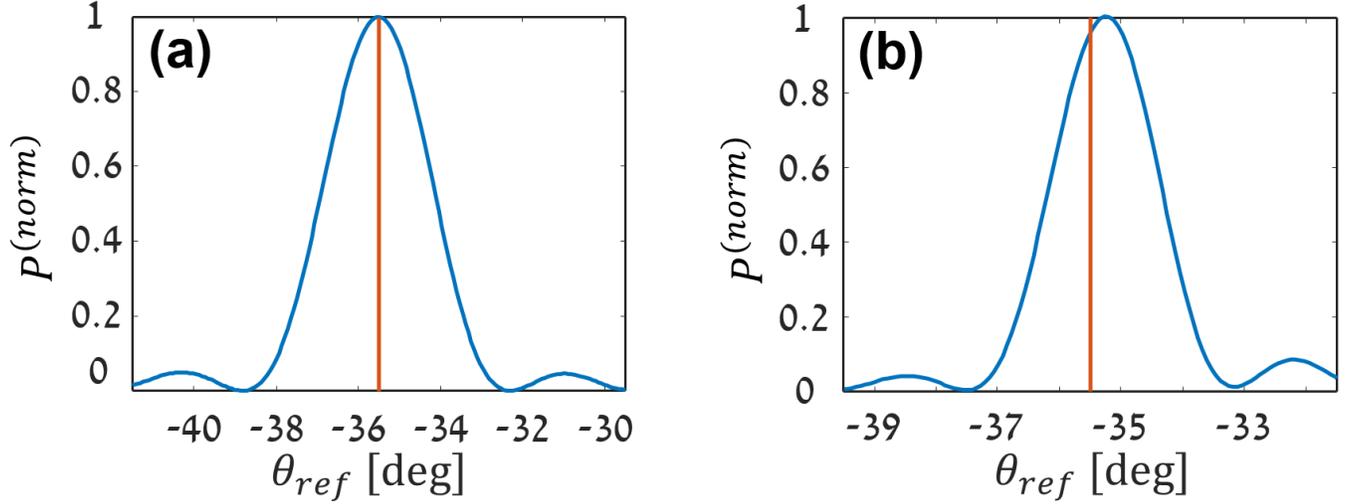

FIG. 7 RA induced reflection anomaly. The normalized power of the specular reflection is presented for the (a) off- and (b) on-resonance cases, according to FIG. 5 (point 3 and point 4 in FIG. 2(b)). The orange vertical lines represent the expected direction of reflection according to $\theta_{ref} = \theta$. While in (a) the reflection is centered around the expected direction, in (b) a shift of $0.25°$ is seen.

Interestingly, the described behavior of the nanoparticle chain is the same for a larger number of inclusions. To see this, we plot in FIG. 6 the trajectory of the same chain as in FIG. 5(d), for the case of $N = 500$ particles. The exact same trend as in FIG. 5(d) can be seen. However, most of the additional particles were added to the spiral at the beginning of the chain. In that regime, the particles roughly respond like they were in an infinite chain. This can be seen by the density of the black arrows (which appear every 10 particles) in the spiral regime vs the monotonic trajectory regime. Thus, the monotonic trajectory regime, where the coherent scattering gradually builds up, occupies a smaller portion of the chain as $N$ increases.

## A. RA Induced Angular Diffraction Anomaly

So far, we have investigated how the RA induces anomalous phase gradients of the particle's excitation. However, the notable implication of these gradients are the associated *angular anomalies* of the light diffracted from the chain to the far-field. Specifically, the direction of specular reflection as well as other diffraction orders from the chain, deviates from the expected direction for infinite chains.

To demonstrate this diffraction anomaly, we compare the angular distribution of the specular reflection of a finite- and an infinite-chain with the same parameters. We first calculate the angle-dependent normalized scattered power by:

$$P^{(norm)}(\theta_{scat}) = \frac{|E(\theta_{scat})|^2}{P_0} \quad (33)$$

$$P_0 = Max(|E(\theta_{scat})|^2)$$

$$E(\theta_{scat}) = \sum_{q=1}^{N} G(|\vec{r}_q - \vec{r}_{scat}|)p_q$$

Where the sum runs over the particles of the chain and $\vec{r}_{scat}$ is set to be in the far-field, pointing to the direction of the specified angle $\theta_{scat}$. In FIG. 7, we show the normalized power of the specular reflection for the two cases shown in FIG. 5 (point 3 and point 4 in FIG. 2(b)). The maximum in Eq. (35) was taken to be from the presented range of $\theta_{ref}$ in the figure, where $\theta_{ref}$ is the angle of reflection. The vertical orange lines show the expected direction of specular reflection, according to $\theta_{ref} = \theta$. We can see that for the off-resonance case, where there is a small range of phase span without a monotonic trend, the reflected power is centered exactly on the expected direction. In contrast, for the on-resonance case, the reflection is shifted by $0.25°$. This reflection anomaly occurs due to the phase gradient of the chain, as seen in FIG. 5(b) and 5(d), and therefore it is also manifested in other diffraction orders. The direction of the observed shift is consistent with the slope of the phase gradient, and its magnitude is equal to 15% of the full width at half the maximum of the specular reflection.

## B. Scattering Path Interference

The presented theoretical framework enables to explore the buildup up the chain response as originating from interference of the different scattering paths. In FIG. 8, we show the scattering paths at the chain, and their interference with each other[3]. The figure presents the various terms that contribute to $b_q$, for the off- (FIG. 8 (a) and FIG. 8(c), $\zeta = 0.05 \cdot e^{0.19i\pi}$) and on- (FIG. 8(b) and FIG. 8(d), $\zeta = 0.37 \cdot e^{0.58i\pi}$) resonance cases.

---

[3] The computation of the partition of $q$ for the different $t_i$ according to $t_1 + 2t_2 + \cdots + qt_q = q$ (Eq. (18)), was performed using [44].

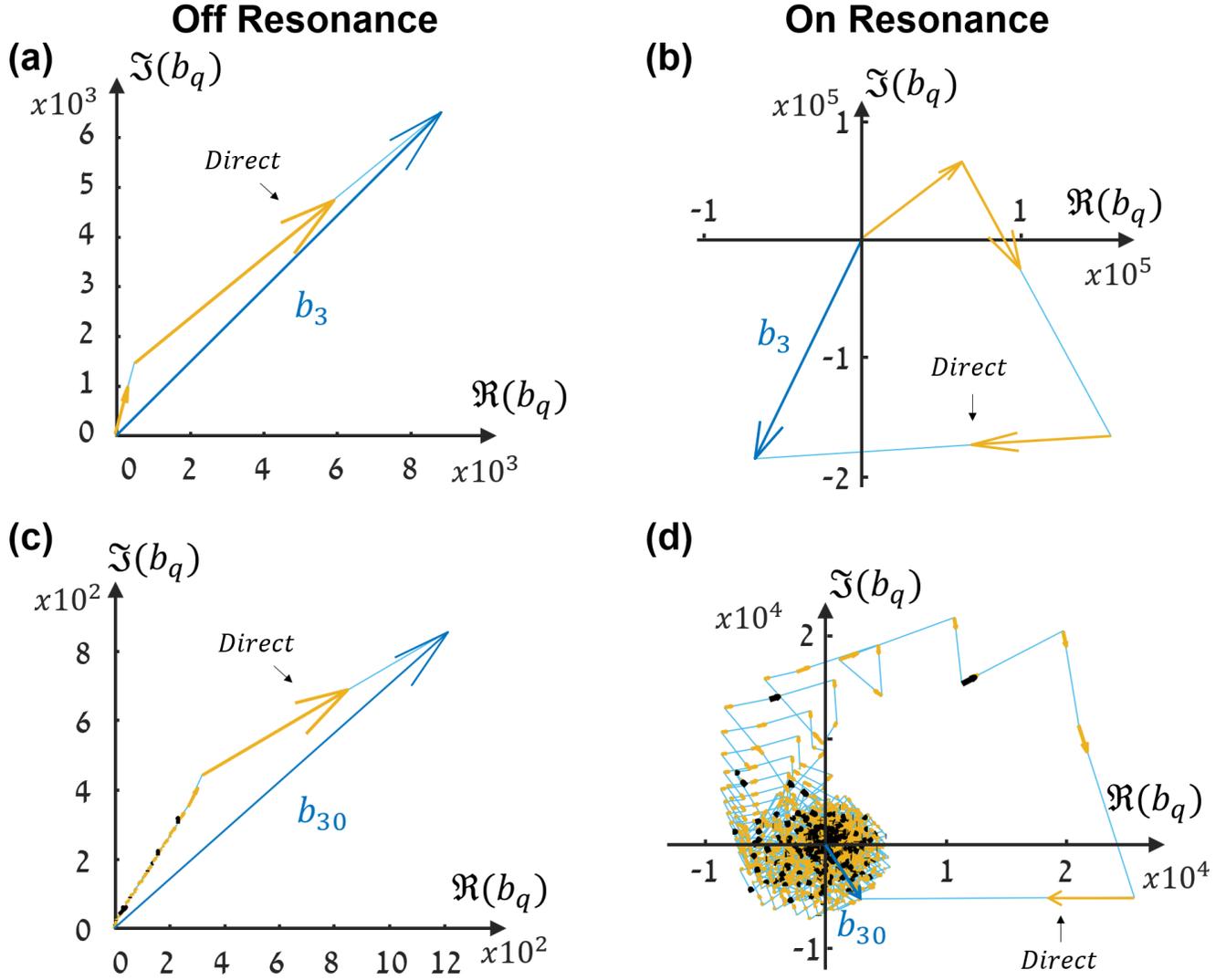

FIG. 8 The interference of the different scattering paths at the chain. The paths contributing to the complex amplitude $b_q$ is shown for $q = 3$ in (a) and (b) and for $q = 30$ in (c) and (d). In addition, (a) and (c) correspond to the off-resonance case (points 1 and 3 in FIG. 2(b)) while (b) and (d) to the on-resonance case (points 2 and 4 in FIG. 2(b)). In (a) and (b) the interference is composed of three scattering paths, according to FIG. 3. We note that the first arrow in (a) is vanishingly small. In the off-resonance case, the different terms add up almost coherently (i.e., with the same phase), giving rise to a total $b_q$ which is considerably larger than each of the paths contributing to it. On the other hand, in the on-resonance case, the path interference is more complex, leading to the observed triangular shapes. This morphology takes place thanks to the resonant response of the nanoparticles, where the phase of $\alpha_s$ is imprinted and cascaded in the interference pattern. Specifically, the paths are ordered by the number of the absorption-remission processes they include (i.e., according to $T \equiv \sum_i t_i$) [44]. Therefore, paths with successive values of $T$ have nearly constant phase difference between them, which gives rise to the observed triangular shapes. The values of $\zeta$ are: Off resonance: $\zeta = 0.05 \cdot e^{0.19 i\pi}$, on resonance: $\zeta = 0.37 \cdot e^{0.58 i\pi}$.

We note that $B_q$ and $b_q$ are intrinsic to the chain, and do not depend on the applied field parameters, nor on the number of particles in the chain. In FIG. 8(a) and 8(b) we show the terms contributing to $b_3$, and in FIG. 8(c) and 8(d), the terms contributing to $b_{30}$. In $b_3$ the interference is composed of three scattering paths, as shown in . The first yellow arrow corresponds to the $1 \to 2 \to 3 \to 4$ path, the second to the $1 \to 2 \to 4$ and $1 \to 3 \to 4$ paths, and the third to the $1 \to 4$ path. We note that the first arrow in FIG. 8(a) is vanishingly small. In the off-resonance case, the different terms add up almost coherently (i.e., with the same phase), giving rise to a total $b_q$ which is considerably larger than each of the paths contributing to it. On the other hand, in the on-resonance case, the path interference is more complex, as the relative angle between successive arrows is significantly larger than in FIG. 9(a). This morphology takes place thanks to the resonant response of the nanoparticles, where the phase of $\alpha_s$ is imprinted and cascaded in the interference pattern. Specifically, the paths are ordered by the number of the absorption-remission processes they include (i.e., according to $T \equiv \sum_i t_i$ [44]. Therefore, paths with successive values of $T$ have nearly constant phase differences between them, which gives rise to the observed triangular shapes.

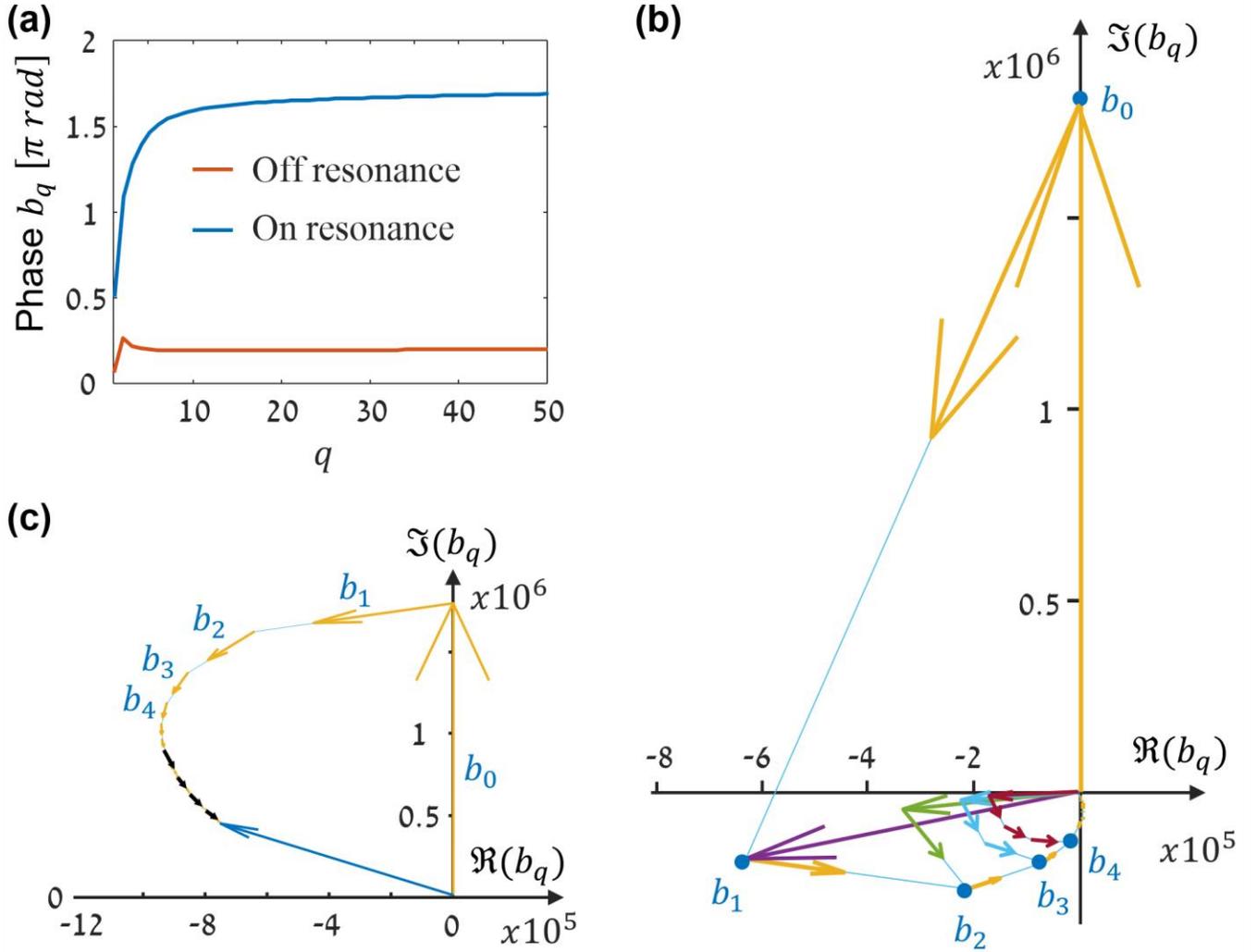

FIG. 9 The dependence of $b_q$ on $q$. (a) The phase of $b_q$ for the off- (points 1 and 3 in FIG. 2(b)) and on- (points 2 and 4 in FIG. 2(b)) resonance cases. In the off-resonance case the phase varies over a small range of less than $0.2\pi$ and saturates very rapidly (after ~5 particles). On the other hand, in the on-resonance case a monotonic phase trend is apparent, spanning more than $\pi$ phase range. In addition, the phase gradient range extends over significantly more particles than in the off-resonance case. (b) The evolution of $b_q$ for the on-resonance case on the complex plane. The blue points represent $b_q$ for different $q$ values. The yellow arrows, with the light blue guiding lines that connect the blue points, represent the difference in $b_q$. The arrows in purple, green, blue and dark red represent the components of the sum in Eq. (34) for $b_1, b_2, b_3$ and $b_4$, respectively. All the different trajectories are curving counterclockwise, according to the positive phase of $\alpha_s$. (c) The trajectory of the accumulative sum of the $b_q$ terms. Each yellow arrow, along with its guiding light blue line, correspond to a certain $b_q$. In addition, every tenth arrow is marked in black ($q = 10, 20, 30, 40$). This trajectory has similar characteristics to the trajectory of $\tilde{p}_q$ from FIG. 5(d).

The mechanism that gives rise to the RA induced phase gradients becomes transparent from the examined behavior of $b_q$. In fact, this phenomenon emerges due to the phase gradient imprinted on $b_q$ as a function of $q$. This gradient origins from the phase of $\alpha_s$ at its resonance, which is cascaded from the chain's edge. To show this, relying on the results of FIG. 8, we present the evolution of $b_q$ as a function of $q$ in FIG. 9. First, in FIG. 9(a) we depict the phase of $b_q$ for the off- (points 1 and 3 in FIG. 2(b)) and on- (points 2 and 4 in FIG. 2(b)) resonance cases. In the off-resonance case the phase varies over a small range of less than $0.2\pi$, where the phase saturates very rapidly (after ~5 particles). This phase trend can be understood from FIG. 8(c), which shows how the different scattering paths add up with almost the same phase. On the other hand, in the on-resonance case, a monotonic phase trend is apparent, spanning more than $\pi$ phase range. In addition, the phase gradient range extends over significantly more particles than in the off-resonance case. To understand the origin of this phase gradient in $b_q$ it is instructive to refer to the derivation according to the generalized Fibonacci series (section II-D). FIG. 9(b) depicts the evolution of $b_q$ for the on-resonance case on the complex plane. The blue points represent $b_q$ for different $q$ values. The yellow arrows, with the light blue guiding lines that connect the blue points, represent the difference in $b_q$. To see how the $b_q$ are recursively generated we can rearrange Eqs. (28)-(29), and write:

$$b_q = \sum_{i=1}^{q} w_i b_{q-i} \tag{34}$$

In FIG. 9(b), we show in purple, green, blue and dark red arrows the components of the sum according to Eq. (34) for $b_1$, $b_2$, $b_3$ and $b_4$, respectively. All the different trajectories are curving counterclockwise, according to the positive phase of $\alpha_s$. In fact, the recursive relation dictates that the initial phase of $b_0$ ($\alpha_s$) will be cascaded onward, which results with a monotonic phase trend. In addition, the first component in the sum ($b_0 w_q$), which corresponds to the direct scattering, has the largest contribution for all $q$ values. Nevertheless, the monotonic phase gradients are attributed to the non-negligible contribution of the remaining sum components. To further connect the analyzed trend of $b_q$ as function of $q$ to the studied response of the chain excitation ($\tilde{p}_q$), we present in FIG. 9(c) the trajectory of the accumulative sum of the $b_q$ terms. Each yellow arrow, along with its guiding light blue line, correspond to a certain $b_q$, as elaborated in the figure. In addition, every tenth arrow is marked in black ($q = 10, 20, 30, 40$). This trajectory has similar characteristics to the trajectory of $\tilde{p}_q$, as seen in FIG. 5(d). By referring to the two-step solution in section II-A and owing to the incoherent scattering towards the right side of the chain, we can roughly approximate $p_{aux}$ as $E_{app}$. Therefore, for a RA towards the left side of the chain only the second step of the solution determines the phase gradients, and the relation $\Delta \tilde{p}_q \approx -b_{N-q}$ holds, as discussed in section II-C. Altogether, we see how the phenomenon of a RA induced phase gradients are encoded in intrinsic characteristic of the chain. The monotonic phase of the $b_q$ terms, determines the monotonic phase of the chain excitation.

### C. RA Induced Transparency

The intriguing effect of an absorption less band induced by coherent scattering has been recently explored [22–24,45,46], showing that attractive slow light features emerge along with an induced transparency. The presented multi-scattering model can be used to explore the spatial evolution of the transparent state. The extinction cross section of the chain can be calculated by summing the extinction of all the nanoparticles [39]:

$$\sigma_{ext} = 4\pi k \sum_{q=1}^{N} \frac{\Im\{E_{app,q}^* p_q\}}{|E_{app,q}|^2} \tag{35}$$

By recalling that $E_{app,q} = E_0 e^{-ik_{\parallel} dq}$, we can simplify this expression to:

$$\sigma_{ext} = \frac{4\pi k}{E_0} \sum_{q=1}^{N} \Im\{\tilde{p}_q\} \tag{36}$$

Therefore, the imaginary part of $\tilde{p}_q$ for each nanoparticle is proportional to its extinction cross section. By reexamining FIG. 5(d), we observe that the extinction of each nanoparticle changes substantially from one edge of the chain to the other. A similar change can also be seen between the single particle response (blue point) to the response of a nanoparticle within an infinite chain (orange point). In case that the parameters of the chain are judiciously chosen, the dipole moment trajectory on the complex plane, as shown in FIG. 5(d), may nearly reach the horizontal axis (zero imaginary part). Hence, the extinction will almost vanish. Interestingly, it means that the chain can become almost entirely transparent, though illuminated at the absorption band of the nanoparticles. The theoretical framework developed in this paper can be beneficial for analyzing effects as this, where the particles' interaction induces a substantial change in light extinction and transmission.

### IV. MACROSCOPIC DESCRIPTION OF THE CHAIN'S DYNAMICS

The theoretical point of view that was adopted at this paper considers the microscopic dynamics of the chain, i.e., analyzes the excitation of each and every particle. Alternatively, the chain dynamics can be viewed by a macroscopic description, via introduction of global wave functions that encapsulate the microscale interactions. In such analysis, the different wave phenomena can be discerned from the analytic properties of the chain's Green function [8]. Specifically, this approach reveals two distinct wave phenomena that are of dominant importance to finite or semi-finite chains with inter-particle spacing larger than $\lambda/2$: leaky modes and continuous spectrum waves. In this case guided modes of any type, confined or light-cone, cannot be supported. Leaky modes can be found by searching the zeros of the chain's equation of dynamics, or equivalently, by seeking for the poles of the chain's spectral Green function, and are characterized by an exponential decay of their amplitude along the chain. Conversely, the continuous spectrum waves are associated with branch cut singularities of the chain's spectral Green's function and shows an algebraic decay [8]. Here we show that the phenomenon discussed in this paper, RA-induced phase gradients, is associated with excitation of a continuous spectrum wave through a diffraction order of the impinging wave wavenumber. The interference between the specular reflection and the resonantly coupled continuous spectrum wave back to the wavenumber of the impinging wave results in the anomaly of the diffracted and reflected light.

The analysis of the chain's modes can be performed by searching for poles of $\alpha_{eff}(\beta)$, defined in Eq. (15), analytically continued to the complex domain. $\beta$ is a complex wave number, which its real part describes the wave periodicity, and its imaginary part describes the associated decay rate. The real part can also be mapped to a corresponding angle of incidence for a plane wave illumination, $\theta_{eff} = \operatorname{asin}(\Re(\beta)/k)$. The dependence of $\alpha_{eff}$ on $\beta$ comes through the analytical continuation of $S(\beta)$, which we can write explicitly for transverse excitation as:

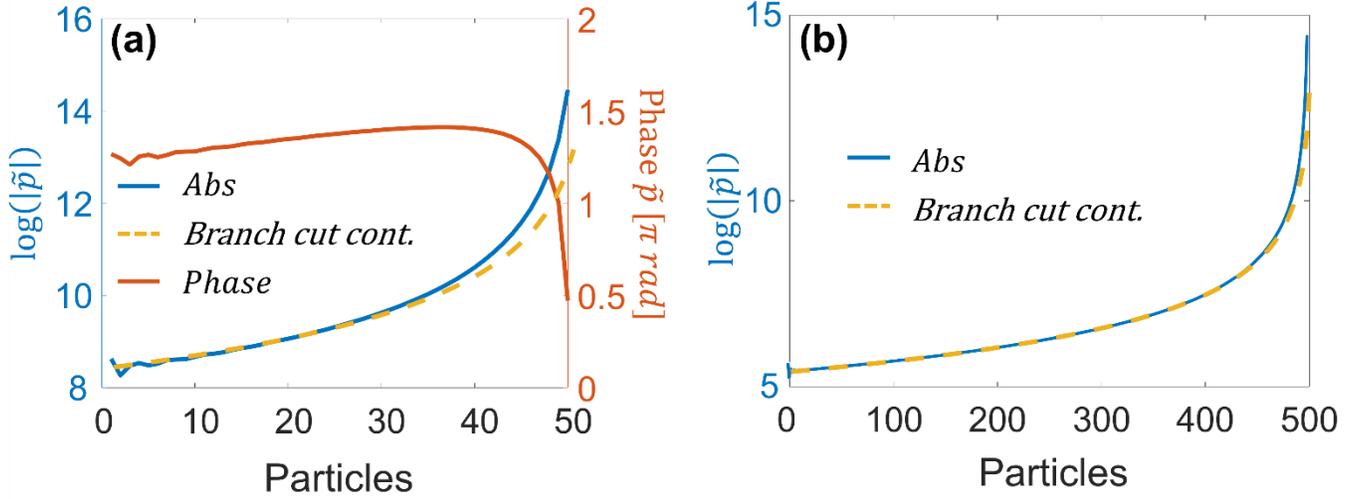

FIG. 10 Continuous spectrum wave excited at the edge of the chain for (a) 50 particles and (b) 500 particles. The figures show the dipole moments along the chain, $\tilde{p} = p/e^{ikx_n}$, for an excitation of only the last particle at the chain's edge. The amplitude (blue solid curve) agrees with the analytic expression for the branch cut contribution of the continues spectrum wave according to Eq. (39) (yellow dashed curve). In addition, the small phase range after the subtraction of the phase associated with $k$ (orange solid curve), confirms the existence of the continuous spectrum wave. The wavelength is $\lambda = 1000\ nm$ and the parameters of the chain are $d = 420\ nm, n = 1.5$ and $\alpha_s = \frac{3i}{2k^3}$.

$$S(\beta) = \sum_{x=-\infty, x\neq 0}^{x=\infty} G_n(x)e^{-i\beta nd}$$
$$= \sum_{n=-\infty, n\neq 0}^{n=\infty} e^{-i\beta nd} e^{ik|n|d} \left[ \frac{k^2}{|n|d} + \frac{ik}{(nd)^2} - \frac{1}{(|n|d)^3} \right] \quad (37)$$
$$= \frac{k^2}{d}\big(Li_1(z_1) + Li_1(z_2)\big)$$
$$+ \frac{ik}{d^2}\big(Li_2(z_1) + Li_2(z_2)\big)$$
$$- \frac{1}{d^3}\big(Li_3(z_1) + Li_3(z_2)\big)$$

Where, $z_1 = e^{id(k+\beta)}$, $z_2 = e^{id(k-\beta)}$, and $Li_s(z)$ is the s$^{\text{th}}$-order Polylogarithm function:

$$Li_s(z) = \sum_{n=1}^{n=\infty} \frac{z^n}{n^s} \quad (38)$$

And the analytical continuation for $|z| > 1$ is defined by $Li_0(z) = z/(1-z)$, $Li_1(z) = -\ln(1-z)$ and $Li_{s+1}(z) = \int_0^z \frac{Li_s(t)}{t} dt$. For $\lambda = 1000\ nm$ and the parameters of the analyzed chain ($d = 420\ nm, n = 1.5, \alpha_s(\lambda = 1000\ nm) = \frac{3i}{2k^3}$) we find a pole of $\alpha_{eff}(\beta)$ at $\beta = (7.4 - 1.8i)\mu m^{-1}$, which corresponds to $\theta_{eff} = $ 52.5° and a decay (of $1/e$) within a distance of $1/(\Im(\beta)d) = 1.28$ particles[4]. This angle and short interaction length confirms that neither a leaky-wave, nor a light-line mode are responsible for the observed behavior. On the other hand, the Polylogarithm function $Li_s(z)$ has a branch cut singularity which starts from the branch point $z = 0$, and extends to infinity on the real axis. This branch points occurs exactly at the RA condition and supports the resonant excitation of a continuous spectrum wave on finite chains. In our case, the resonant response of the nanoparticles stimulates the resonant excitation of the continuous spectrum wave, which decays over a long distance along the chain length. To confirm the existence of the discussed wave phenomenon we examine, in FIG. 10, the chain's response for an excitation of only a single particle at its edge. The continuous spectrum wave is expected to be excited and match a periodicity according to $k$ (wavenumber in the medium surrounding the linear chain), with an asymptotic decay (for large number of particles) of the form [8]:

$$G_n^{cw} = \frac{d}{k^2} \frac{e^{i|n|kd}}{|n|} \prod_{m=1}^{2} \frac{1}{\ln|n| + A_m} \quad (39)$$

where $n$ is the particle numeration from the excited edge, $A_{1,2} = C \pm i\pi$, $C = -\frac{kd}{\alpha_s} + Li_1(e^{2ikd}) + \frac{i}{kd}\left(Li_2(e^{2ikd}) + \frac{\pi^2}{6}\right) - \frac{1}{(kd)^2}\left(Li_3(e^{2ikd}) + \zeta(3)\right)$ and $\zeta(3)$ is the Riemann zeta function of 3. Eq. (39) accounts only for the branch cut contribution of the chain's response. In FIG. 10, we show the dipole moment excitations on a logarithmic scale (blue solid line) for a chain of 50 particles (a) and 500 particles (b). First, the nonlinear trend of the curve confirms again that neither a leaky-wave nor a light-line mode are dominant in the observed trend. In addition, the dashed yellow curve shows the function $G_n^{cw}$ according to Eq. (39). Very good agreement is seen as $n$ gets larger, as expected for asymptotic approximation. The slight deviations at the end of the chain are associated with reflections from the chain's edge. Additionally, we show the phase of the dipole moments, where we have subtracted the phase associated with $k$: $\tilde{p} = p/e^{ikx_n}$. We see that the wave periodicity agrees with the wavenumber k. Overall, we see the connection between the microscopic analysis for the origin of the RA-induced phase gradient, to the macroscopic description, which reveals excitation of a continuous spectrum wave.

---

[4] We note that this pole exists in the zeroth Riemann sheet for the Polylogarithm functions that depend on $z_1$ and in the first Riemann sheet for those who depend on $z_2$.

## V.     CONCLUSIONS

To summarize, in this paper, we developed a combinatorial multiple scattering model to analyze the spatial excitation in finite nanoparticle chains. With this model, we analyzed a novel phenomenon of a RA induced phase gradients in finite nanoparticle chains. These phase gradients are inherent to finite chains with a resonant response of the nanoparticles and in spectral vicinity to a unidirectional RA, i.e., towards one direction of the chain. The induced monotonic phase profiles also affect the diffraction and specular reflection from the chain. Specifically, it causes a deviation of the diffraction orders relative to the direction expected for an infinite chain.

The analytical model treats the chain, a reciprocal system, by a successive solution of two non-reciprocal systems. Then, by approximating the governing matrices, we find an approximated close form solution to the particle dipole moments. Intriguingly, the solution enables describing the excitation as originating from the interference of multiple scattering paths within the chain. Moreover, we show that the multiple scattering process can be accounted for by the generalized Fibonacci series, which reveals the recursive relation that underlies the chain excitation. The model presented in this paper is important for characterizing and analyzing finite nanoparticle arrays and the various phenomena that they can exhibit. Specifically, the analysis of leaky-wave radiation and beam shaping can benefit from this treatment. Moreover, the model may find use also in the description of finite disordered systems or different systems with weakly coupled components.

**Acknowledgments**
This publication is part of a project that has received funding from the European Research Council (ERC) under the European Union's Horizon 2020 research and innovation programme (grant agreement No 715362 and grant agreement No 639402), and by the Israel Science Foundation (grant agreement 581/19) (grant agreement No 3-15614).

\* *liormic1@caltech.edu*

# Supplemental Material for
# Rayleigh Anomaly Induced Phase Gradients in Finite Nanoparticle Chains

Lior Michaeli[1-4,*], Ofer Doron[1-3], Yakir Hadad[1], Haim Suchowski[2,3], Tal Ellenbogen[1,3]

[1] *Department of Physical Electronics, Faculty of Engineering, Tel-Aviv University, Tel-Aviv 6779801, Israel*
[2] *Raymond and Beverly Sackler School of Physics & Astronomy, Tel-Aviv University, Tel-Aviv 6779801, Israel*
[3] *Center for Light-Matter Interaction, Tel-Aviv University, Tel-Aviv 6779801, Israel*
[4] Currently at: *Thomas J. Watson Laboratories of Applied Physics, California Institute of Technology, Pasadena, California, 91125, USA*

## 1. Analytical condition for model accuracy:

In order to analyze the accuracy of the model approximation, presented in Eq. (12) of the main text, we will rely on the theory of sensitivity analysis in linear systems [1]. For the linear system $\hat{A}p = E_{app}$, any perturbation of $\hat{A}$ to $\hat{A} + \Delta\hat{A}$ will result in a corresponding perturbation of the solution vector $p$:

$$(\hat{A} + \Delta\hat{A})(p + \Delta p) = E_{app} \tag{S1}$$

Quantification of the solution's error is given by means of its norm [1]:

$$\frac{\|\Delta p\|}{\|p\|} \leq cond(\hat{A})\frac{\|\Delta\hat{A}\|}{\|\hat{A}\|} \tag{S2}$$

Where $\|p\|$ denotes the norm of the vector $p$, $\|A\|$ and $\|\Delta\hat{A}\|$ denote the norm of the corresponding matrices, and $cond(\hat{A})$ is the condition number of the matrix $\hat{A}$, which determines the sensitivity of the solution to small changes in the input data. In order to quantify the expression in Eq. S2, we first write an exact expression to $\|\hat{A}\|$, then we derive an upper bound to $\|\Delta\hat{A}\|$. For that, we will use the definition of a maximum norm to the matrices: $\|\hat{A}\|_M = \max_{i,j}|A_{ij}|$. In addition, in order to simplify the derivation, we will account only for the radiative term in the green function $g_q$ (Eq. (3)), and write $g_q = \frac{g_1}{q}$. This approximation is justified as long as near fields do not play an important role in the interaction between the particles.

Calculation of $\|\hat{A}\|$:

The magnitude of largest element of $\hat{A}$ is the value of the diagonal elements:

$$\|\hat{A}\|_M = |A_{ii}|^2 = |\alpha_s|^{-2} \tag{S3}$$

Upper bound to $\|\Delta\hat{A}\|$:

To estimate $\|\Delta\hat{A}\|$ we first need to obtain $\Delta\hat{A}$. Using the definition $\Delta\hat{A} \equiv \hat{A}_{Model} - \hat{A}$, we can perform the explicit multiplication to find $\Delta\hat{A}$. For example, for $N = 4$ we get:

$$\Delta \hat{A} = -\alpha_s \begin{bmatrix} 0 & 0 & 0 & 0 \\ 0 & g_1^2 & g_1 g_2 & g_1 g_3 \\ 0 & g_1 g_2 & g_1^2 + g_2^2 & g_1 g_2 + g_2 g_3 \\ 0 & g_1 g_3 & g_1 g_2 + g_2 g_3 & g_1^2 + g_2^2 + g_3^2 \end{bmatrix} \approx -\alpha_s g_1^2 \begin{bmatrix} 0 & 0 & 0 & 0 \\ 0 & \frac{1}{1} & \frac{1}{1} \cdot \frac{1}{2} & \frac{1}{1} \cdot \frac{1}{3} \\ 0 & \frac{1}{1} \cdot \frac{1}{2} & \frac{1}{1^2} + \frac{1}{2^2} & \frac{1}{1} \cdot \frac{1}{2} + \frac{1}{2} \cdot \frac{1}{3} \\ 0 & \frac{1}{1} \cdot \frac{1}{3} & \frac{1}{1} \cdot \frac{1}{2} + \frac{1}{2} \cdot \frac{1}{3} & \frac{1}{1^2} + \frac{1}{2^2} + \frac{1}{3^2} \end{bmatrix} \quad (S4)$$

The largest element is the last element on the diagonal:

$$\|\Delta \hat{A}\|_\infty = |\alpha_s|^2 |g_1|^4 \left| \sum_{i=1}^{N-1} \frac{1}{i^2} \right|^2 \leq |\alpha_s|^2 |g_1|^4 \frac{\pi^2}{6} \quad (S5)$$

Where in the last step we used $\lim_{N \to \infty} \sum_{i=1}^{N-1} \frac{1}{i^2} = \pi^2/6$ to write the upper bound.

Mathematical condition for the model validity:

Using Eqs. (S2), (S3) and (S5) we can now obtain the condition for the validity of the model by demanding $cond(A) \ll \frac{\|\hat{A}\|}{\|\Delta \hat{A}\|}$:

$$cond(A) \ll \frac{6}{\pi^2} \cdot \frac{1}{|\zeta|^4} \equiv C(\zeta) \quad (S6)$$

This condition was derived without accounting for the specific illumination field $E_{app}$, and therefore is true for any arbitrary illumination profile. We also note that this is a stringent condition, as we have taken a lower bound to $\|\hat{A}\|/\|\Delta \hat{A}\|$. The derived expression $C(\zeta)$ strongly diverges as $|\zeta|$ goes to zero, showing that for weakly interacting particles the model is exact. In order to estimate the validity of the model for larger $|\zeta|$ the condition number of the interaction matrix $\hat{A}$ should be evaluated. For all the parameters considered at the manuscript, the condition number calculated according to the maximum norm definition $cond(\hat{A}) = \|\hat{A}\|_M \|\hat{A}^{-1}\|_M$ is on the order of $cond(\hat{A}) \approx 1$. Figure S2 shows the dependence of $C(\zeta)$ on $|\zeta|$. As the curve of $C(\zeta)$ gets larger than one the model becomes more accurate.

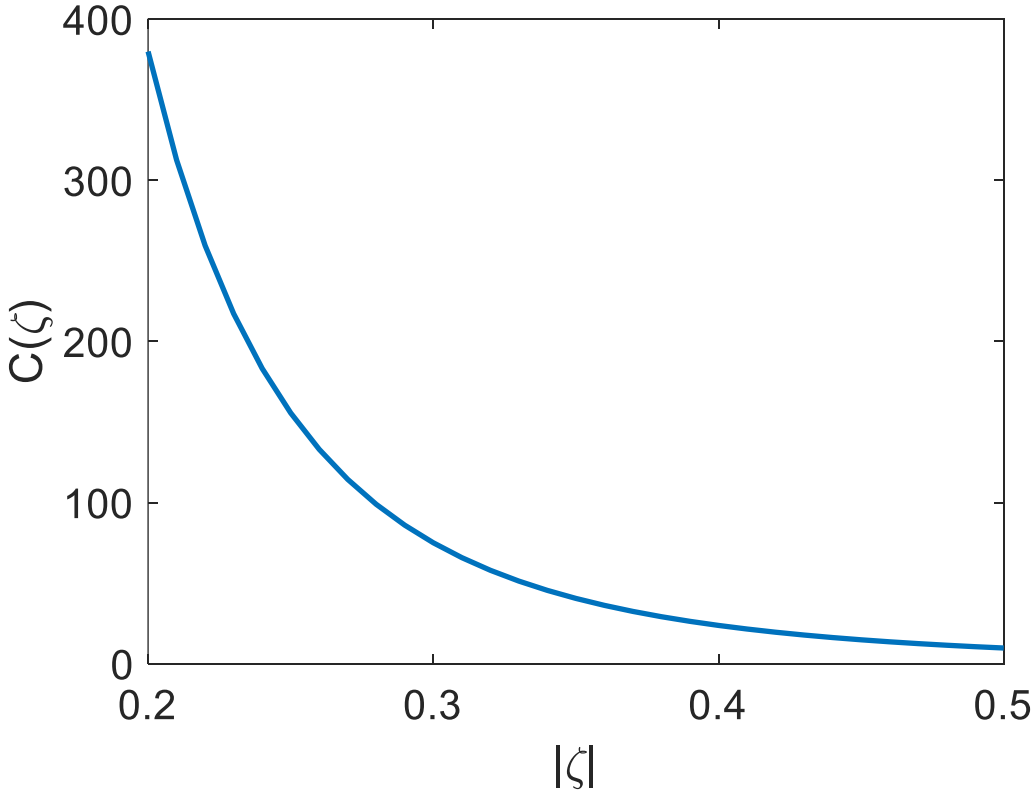

FIG. S2 The dependence of the model accuracy on $|\zeta|$. For accurate results the interaction matrix $\hat{A}$ should satisfy $cond(A) \ll C(\zeta)$.

## 2. Numerical verification of model accuracy:

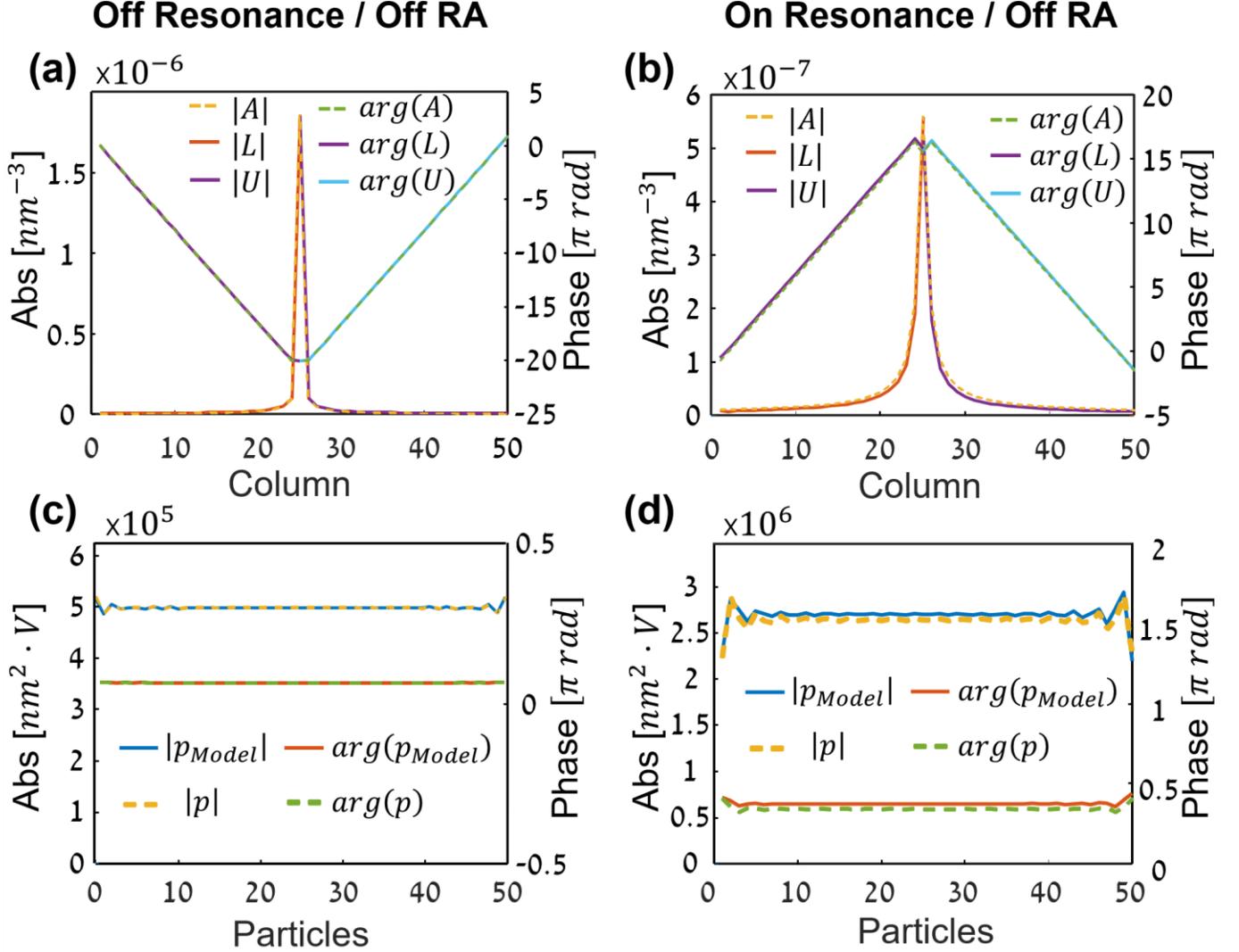

FIG. S1 Agreement of the model and the exact solution. Comparison of the amplitude and phase of the matrices $L$ and $U$ vs matrix $A$, for the case of off (a) and on (b) resonance, both far from a RA. These cases correspond to point 1 and point 2 in FIG. 2**Error! Reference source not found.**(b), respectively. The comparison is for the middle (25$^{th}$) row. The associated particles' dipole moments solved based on the derived model (Eqs. **Error! Reference source not found.** and **Error! Reference source not found.**) (solid lines) vs the exact solution (dashed lines) is shown in (c) and (d), correspondingly. The parameters of the two cases are: Off resonance - $|\alpha_s| = 0.3|\alpha_s^{Max}|, arg(\alpha_s) = 0.07\pi, \lambda = 1440\ nm$, and on resonance - $|\alpha_s| = |\alpha_s^{Max}|, arg(\alpha_s) = 0.5\pi, \lambda = 1000\ nm$. The comparison is for a 50-particle chain with spacing $d = 420\ nm$ and surrounding refractive index $n = 1.5$.

In FIG. S3(a) and S3(b), we directly examine the model assumption in Eq. **Error! Reference source not found.**, by comparing cross sections of the amplitude and phase of the three matrices $\hat{A}, \hat{L}$ and $\hat{U}$ for two cases: First case (point 1 in FIG. 2**Error! Reference source not found.**(b)) - $|\alpha_s| = 0.3|\alpha_s^{Max}|, arg(\alpha_s) = 0.07\pi$ and $\lambda = 1440\ nm$. Second case (point 2 in FIG. 2**Error! Reference source not found.**(b)) - $|\alpha_s| = |\alpha_s^{Max}|, arg(\alpha_s) = 0.5\pi$ and $\lambda = 1000\ nm$. The comparison is for a chain with $N = 50$ particles. Very good agreement for both cases is seen. In FIG. S3(c) and S3(d) we present the associated particles' dipole moments solved based on the derived model (Eqs. **Error! Reference source not found.** and **Error! Reference source not found.**) (solid lines) vs the exact solution (dashed lines). Also here, both cases show excellent agreement between the model and the exact solution. The two cases shown in FIG. S3 correspond to $\zeta = 0.05 \cdot e^{0.19i\pi}$ (off resonance) and $\zeta = 0.37 \cdot e^{0.58i\pi}$ (on resonance).

## 3. Scattered Fields:

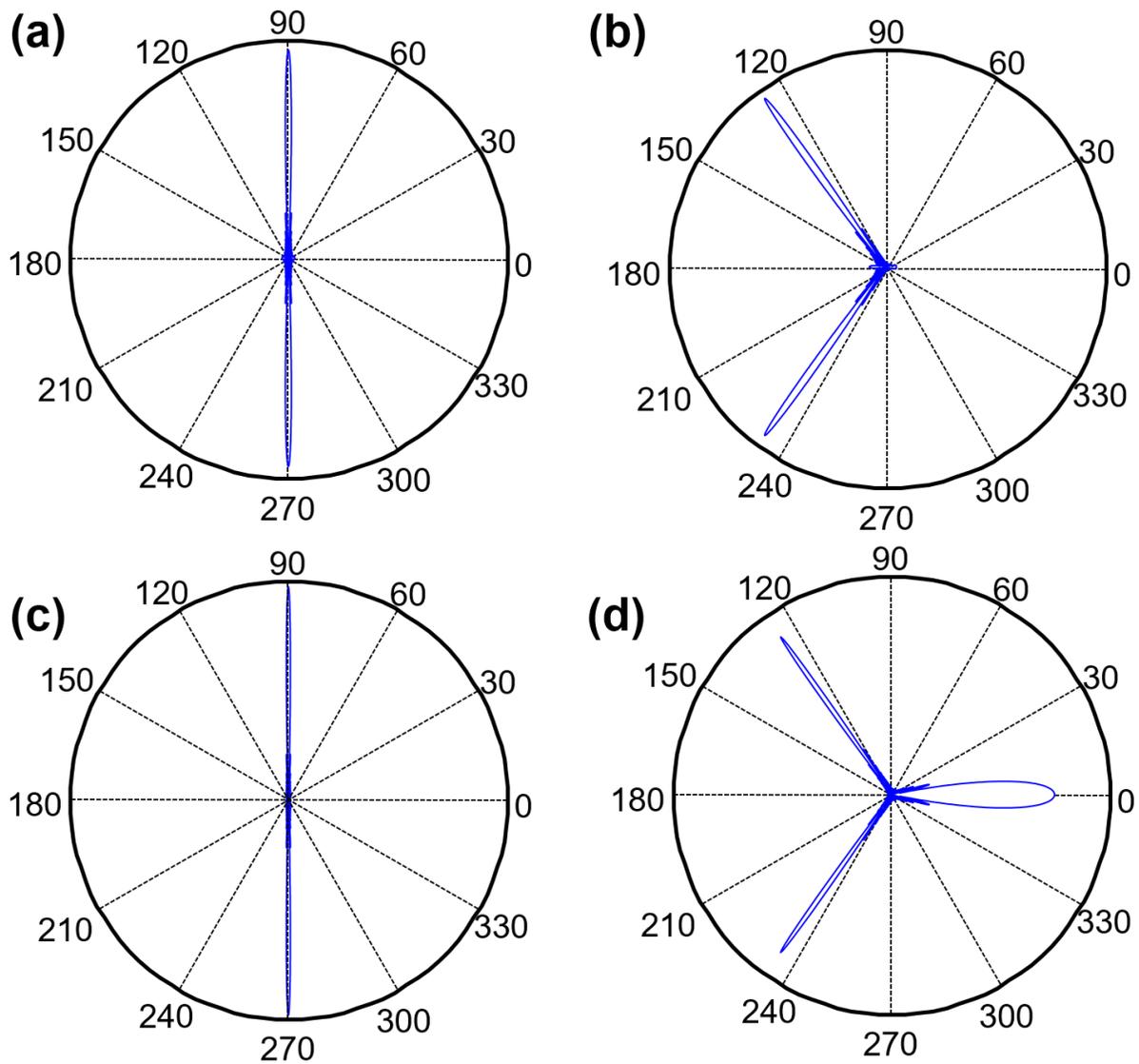

FIG. S1 The scattered fields' polar plots associated with the 4 points shown in Fig. 2: (a) - point 1, (b) – point 3, (c) – point 2, (d) – point 4.